\documentclass[11pt]{article}
\usepackage{amstext}
\usepackage{amsmath}
\usepackage{graphicx}
\usepackage{url}
\usepackage{geometry}
\title{Estimating Animal Abundance with N-Mixture Models Using the \textsf{R-INLA} Package for \textsf{R}}
\author{Timothy D. Meehan, Nicole L. Michel\\National Audubon Society \\ \\
  and H{\aa}vard Rue\\King Abdulla University of Science and Technology}

\begin{document}
\maketitle
\renewcommand{\abstractname}{\vspace{-\baselineskip}}

\begin{abstract}
\normalsize
Successful management of wildlife populations requires accurate estimates of abundance. Abundance estimates can be confounded by imperfect detection during wildlife surveys. N-mixture models enable quantification of detection probability and, under appropriate conditions, produce abundance estimates that are less biased. Here, we demonstrate use of the \textsf{R-INLA} package for \textsf{R} to analyze N-mixture models and compare performance of \textsf{R-INLA} to two other common approaches: \textsf{JAGS} (via the \textsf{runjags} package for \textsf{R}), which uses Markov chain Monte Carlo and allows Bayesian inference, and the \textsf{unmarked} package for \textsf{R}, which uses maximum likelihood and allows frequentist inference. We show that \textsf{R-INLA} is an attractive option for analyzing N-mixture models when (\textit{i}) fast computing times are necessary (\textsf{R-INLA} is 10 times faster than \textsf{unmarked} and 500 times faster than \textsf{JAGS}), (\textit{ii}) familiar model syntax and data format (relative to other \textsf{R} packages) is desired, (\textit{iii}) survey-level covariates of detection are not essential, and (\textit{iv}) Bayesian inference is preferred.
\end{abstract}

\section{Introduction}
\subsection{Background}
Successful management of wildlife species requires accurate estimates of abundance \cite{Yoccoz_Nichols_Boulinier_2001}. One common method for estimating animal abundance is direct counts \cite{Pollock_Nichols_Simons_Farnsworth_Bailey_Sauer_2002}. Efforts to obtain accurate abundance estimates via direct counts can be hindered by the cryptic nature of many wildlife species, and by other factors such as observer expertise, weather, and habitat structure \cite{Denes_Silveira_Beissinger_2015}. The lack of perfect detection in wildlife surveys is common, and can cause abundance to be underestimated \cite{Joseph_Elkin_Martin_Possingham_2009}.

In recent years, new survey designs and modeling approaches have enabled improved estimates of animal abundance that are less biased by imperfect detection \cite{Denes_Silveira_Beissinger_2015}. One such survey design, termed a metapopulation design \cite{Kery_Royle_2010}, involves repeat visits in rapid succession to each of multiple study sites in a study area. If, during repeat visits, the population is assumed to be closed (no immigration, emigration, reproduction or mortality; i.e., static abundance), then information on detections and non-detections during repeated counts can inform an estimate of detection probability. This detection probability can be used to correct abundance estimates for imperfect detection \cite{Royle_2004}.

Data resulting from this survey design are often modeled using an explicitly hierarchical statistical model referred to in the quantitative wildlife ecology literature as an N-mixture model \cite{Royle_Nichols_2003, Dodd_Dorazio_2004, Royle_2004, Kery_Royle_Schmid_2005}. One form of an N-mixture model, a binomial mixture model, describes individual observed counts $y$ at site $i$ during survey $j$ as coming from a binomial distribution with parameters for abundance $N$ and detection probability $p$, where $N$ per site is drawn from a Poisson distribution with an expected value $\lambda_i$. Specifically,

$$N_i \sim \text{Pois}(\lambda_i) \qquad \text{and} \qquad  y_{i,j} | N_i \sim \text{Bin}(N_i, p_{i,j}).$$

$\lambda$ is commonly modeled as a log-linear function of site covariates, as $\text{log}(\lambda_i) = \beta_0 + \beta_1 x_i$. Similarly, $p$ is commonly modeled as $\text{logit}(p_{i,j}) = \alpha_0 + \alpha_1 x_{i,j}$, a logit-linear function of site-survey covariates.

This estimation approach can be extended to cover $K$ distinct breeding or wintering seasons, which correspond with distinct years for wildlife species that are resident during annual breeding or wintering stages \cite{Kery_Dorazio_Soldaat_Van_Strien_Zuiderwijk_Royle_2009}. In this case, population closure is assumed across $J$ surveys within year $k$, but is relaxed across years \cite{Kery_Dorazio_Soldaat_Van_Strien_Zuiderwijk_Royle_2009}. A simple specification of a multiple-year model is $N_{i,k} \sim \text{Pois}(\lambda_{i,k}), \ y_{i,j,k} | N_{i,k} \sim \text{Bin}(N_{i,k}, p_{i,j,k})$. Like the single-year specification, $\lambda$ is commonly modeled using site and site-year covariates, and $p$ using site-survey-year covariates. 

There are other variations of N-mixture models that accommodate overdispersed counts through use of a negative binomial distribution \cite{Kery_Royle_2010}, a zero-inflated Poisson distribution \cite{Wenger_Freeman_2008}, or survey-level random effects \cite{Kery_Schaub_2011}, or underdispersed counts using mixtures of binomial and Conway-Maxwell-Poisson distributions \cite{wu2015bayesian}. Yet other variations account for non-independent detection probabilities through use of a beta-binomial distribution  \cite{Martin_Royle_Mackenzie_Edwards_Kery_Gardner_2011}, parse different components of detection through the use of unique covariates  \cite{O'Donnell_Thompson_III_Semlitsch_2015}, or relax assumptions of population closure  \cite{Chandler_Royle_King_2011, Dail_Madsen_2011}. We do not discuss all of these variations here, but refer interested readers to \cite{Denes_Silveira_Beissinger_2015} for an overview, and to  \cite{Barker_Schofield_Link_Sauer_et_al_2017} for a discussion of assumptions and limitations.

The development of metapopulation designs and N-mixture models represents a significant advance in quantitative wildlife ecology. However, there are practical issues that sometimes act as barriers to adoption. Many of the examples of N-mixture models in the wildlife literature have employed Bayesian modeling software such as \textsf{WinBUGS}, \textsf{OpenBUGS}, \textsf{JAGS}, or \textsf{Stan}  \cite{plummer2003jags,Lunn_Jackson_Best_Thomas_Spiegelhalter_2012,Carpenter_Gelman_Hoffman_Lee_et_al_2017}. These are extremely powerful and flexible platforms for analyzing hierarchical models, but they come with a few important challenges. First, many wildlife biologists are not accustomed to coding statistical models using the \textsf{BUGS} or \textsf{Stan} modeling syntax. While there are several outstanding resources aimed at teaching these skills \cite{Royle_Dorazio_2008, Kery_2010, Kery_Schaub_2011, Kery_Royle_2015, Korner-Nievergelt_Roth_et_al_2015} learning them is, nonetheless, a considerable commitment. Second, while Markov chain Monte Carlo (MCMC) chains converge quickly for relatively simple N-mixture models, convergence for more complex models can take hours to days, or may not occur at all \cite{Kery_Schaub_2011}.

There are other tools available for analyzing N-mixture models that alleviate some of these practical issues. The \textsf{unmarked} package \cite{Fiske_Chandler_2011} for \textsf{R} statistical computing software \cite{R_Core_Team_2016} offers several options for analyzing N-mixture models within a maximum likelihood (ML) framework, with the capacity to accommodate overdispersed counts and dynamic populations. The model coding syntax used in \textsf{unmarked} is a simple extension of the standard \textsf{R} modeling syntax. Models are analyzed using ML, so model analysis is often completed in a fraction of the time taken using MCMC. The familiar model syntax and rapid model evaluation of \textsf{unmarked} has undoubtedly contributed to the broader adoption of N-mixture models by wildlife biologists. However, it comes at a cost, loss of the intuitive inferential framework associated with Bayesian analysis.

Here we demonstrate analysis of N-mixture models using the \textsf{R-INLA} package  \cite{Martins_Simpson_Lindgren_Rue_2013,Rue_Riebler_Sorbye_Illian_Simpson_Lindgren_2017} for \textsf{R}. The \textsf{R-INLA} package uses integrated nested Laplace approximation (INLA) to derive posterior distributions for a large class of Bayesian statistical models that can be formulated as latent Gaussian models \cite{Rue_Martino_Chopin_2009, Lindgren_Rue_Lindstrom_2011}. INLA was developed to allow estimation of posterior distributions in a fraction of the time taken by MCMC. Like \textsf{unmarked}, the model syntax used by the \textsf{R-INLA} package is a straightforward extension of the modeling syntax commonly used in \textsf{R}. Also, like \textsf{unmarked}, the computational cost of analyzing models with \textsf{R-INLA} is relatively low compared to MCMC. The \textsf{R-INLA} approach is different from \textsf{unmarked} in that inference about model parameters falls within a Bayesian framework.

\subsection{Overall objectives}
The purpose of this manuscript is to present a comparative analysis of N-mixture models that is centered on the \textsf{R-INLA} package. In the process, we employ both simulated and real count datasets, and analyze them using \textsf{R-INLA}, \textsf{JAGS}, via the \textsf{runjags} package \cite{Denwood_2016} for \textsf{R}, and the \textsf{unmarked} package for \textsf{R}. In each case, we demonstrate how models are specified, how model estimates compare to simulation inputs and to each other, and how methods compare in terms of computational performance. When describing \textsf{R-INLA} analyses, we detail the format of input data and the content of analysis code, to facilitate readers conducting their own analyses.

We also explore a limitation of the \textsf{R-INLA} approach related to model specification. In particular, while it is possible to specify survey-level covariates for detection using \textsf{JAGS} and \textsf{unmarked}, this is not possible using \textsf{R-INLA}. Rather, survey-level covariates of detection must be averaged to the site or site-year level. Using an averaged detection covariate does allow accounting for site or site-year differences in survey conditions, should they occur. However, in the process of averaging, information related to detection within a site or site-year combination is discarded, which could lead to biased detection and abundance estimates under certain conditions.

Much of the code used to conduct the \textsf{R-INLA} analyses is shown in the body of this manuscript. However, some repeated \textsf{R-INLA} code, code used in \textsf{JAGS} and \textsf{unmarked} analyses, and code related to generating figures, is not shown, for brevity. All code, fully commented, can be accessed via \url{https://github.com/tmeeha/inlaNMix}. Regarding code, note that the \textsf{R-INLA} package is atypical among \textsf{R} packages in a few different ways.  First, \textsf{R-INLA} is not available on the Comprehensive R Archive Network (CRAN), as are many other \textsf{R} packages. Second, \textsf{R-INLA} was initially called \textsf{INLA}, based on its origin as a stand-alone \textsf{C} program.  Over time, community reference to the packaged evolved to become \textsf{R-INLA}.  However, installing and loading the package still employs the original name, which may cause some confusion. To install the package, paste \textsf{install.packages("INLA", repos="https://inla.r-inla-download.org/R/stable")} into an \textsf{R} console. To load the package, use the \textsf{R} command \textsf{library(INLA)}. See \url{https://r-inla.org} to connect with the community around the development of \textsf{R-INLA} and its application to geostatistics, biostatistics, epidemiology, and econometrics \cite{Lindgren_Rue_2015,Blangiardo_Cameletti_2015}.

\section{Example data}
\subsection{Simulated data}
The data simulated for Example I (Section 3) and Example II (Section 4) were intended to represent a typical wildlife abundance study. To put the simulation into context, consider an effort to estimate the abundance of a bird species in a national park, within which are located 72 study sites. At each site, 3 replicate surveys are conducted within 6 weeks, during the peak of the breeding season, when birds are most likely to be singing. In order to estimate a trend in abundance over time, clusters of repeated surveys are conducted each breeding season over a 9-year period.

In this scenario, the abundance of the species is thought to vary with two site-level covariates (\textit{x1} and \textit{x2}), which represent habitat characteristics at a site and do not change appreciably over time, and a third covariate that indicates the year (\textit{x3}). The detection probability is believed to vary according to two covariates (\textit{x1} and \textit{x4}). The first covariate for detection, \textit{x1}, is the same site-level \textit{x1} that affects abundance, although it has the opposite effect on detection. The other detection covariate, \textit{x4}, is a site-survey-year variable that could be related to weather conditions during an individual survey. As is common, due to effects of unknown variables, simulated counts were overdispersed. Overdispersed counts were generated and modeled using a negative binomial distribution. Simulation data was generated using the model

$$N_{i,k} \sim \text{NegBin}(\lambda_{i,k}, \theta) \qquad \text{and} \qquad  y_{i,j,k} | N_{i,k} \sim \text{Bin}(N_{i,k}, p_{i,j,k}),$$

where $\lambda$ was a log-linear function of site and year covariates, as $\text{log}(\lambda_{i,k}) = \beta_0 + \beta_1 (\textit{x1}_{i}) + \beta_2 (\textit{x2}_{i}) + \beta_3 (\textit{x3}_{k})$. $p$ was a logit-linear function of site and site-survey-year covariates, as $\text{logit}(p_{i,j,k}) = \alpha_0 + \alpha_1 (\textit{x1}_{i}) + \alpha_4 (\textit{x4}_{i,j,k})$.

Parameter values for the linear predictor for $\lambda$ were set to $\beta_0$ = 2.0, $\beta_1$ = 2.0, $\beta_2$ = -3.0, $\beta_3$ = 1.0. The overdispersion parameter was set to $\theta$ = 3.0. Parameter values for the linear predictor for $p$ were set to: $\alpha_0$ = 1.0, $\alpha_1$ = -2.0, $\alpha_4$ = 1.0. All independent variables in the simulation were centered at zero to reduce computational difficulties and to make model intercepts more easily interpreted.

We simulated data for Examples I and II using the \texttt{sim.nmix()} function, shown below, with which we encourage readers to experiment. Parameter and variable names in the function code are similar to those given in the model description, above. Note that the function produces two versions of detection covariate \textit{x4} (\texttt{x4} and \texttt{x4.m}) and two versions of the count matrix (\texttt{Y} and \texttt{Y.m}). Covariate \texttt{x4} is the same as the site-survey-year variable \textit{x4}, described above. It is used to generate \texttt{Y}, which is used in Example II. Covariate \texttt{x4.m} is derived from \texttt{x4}, where values are unique to site and year, but are averaged and duplicated over surveys. It is used to generate \texttt{Y.m}, which is employed in Example I. Running \texttt{sim.nmix()} results in a list containing data frames for use with \textsf{R-INLA} and \textsf{unmarked}, and values and vectors for use with \textsf{JAGS}. Before running the function, we install and load libraries and set the seed for the random number generator so that the results are reproducible.

\begin{verbatim}
R> install.packages("INLA", repos="https://inla.r-inla-download.org/R/stable")
R> library(INLA)
R> install.packages(c("runjags", "unmarked"))
R> library(runjags)
R> library(unmarked)
R> set.seed(12345)
R> sim.nmix <- function(n.sites = 72,    # number of study sites
+    n.surveys = 3,                      # short term replicates
+    n.years = 9,                        # number of years
+    b0 =  2.0,                          # intercept log(lambda)
+    b1 =  2.0,                          # x1 slope log(lambda)
+    b2 = -3.0,                          # x2 slope log(lambda)
+    b3 =  1.0,                          # x3 slope log(lambda)
+    a0 =  1.0,                          # intercept logit(p)
+    a1 = -2.0,                          # x1 slope logit(p)
+    a4 =  1.0,                          # x4 slope logit(p)
+    th =  3.0                           # overdisperison parameter
+    ){
+
+    # make empty N and Y arrays
+    if(n.years %% 2 == 0) {n.years <- n.years + 1}
+    N.tr <- array(dim = c(n.sites, n.years))
+    Y <- array(dim = c(n.sites, n.surveys, n.years))
+    Y.m <- array(dim = c(n.sites, n.surveys, n.years))
+
+    # create abundance covariate values
+    x1 <- array(as.numeric(scale(runif(n = n.sites, -0.5, 0.5), scale = F)),
+      dim = c(n.sites, n.years))
+    x2 <- array(as.numeric(scale(runif(n = n.sites, -0.5, 0.5), scale = F)),
+      dim = c(n.sites, n.years))
+    yrs <- 1:n.years; yrs <- (yrs - mean(yrs)) / (max(yrs - mean(yrs))) / 2
+    x3 <- array(rep(yrs, each = n.sites), dim = c(n.sites, n.years))
+
+    # fill true N array
+    lam.tr <- exp(b0 + b1 * x1 + b2 * x2 + b3 * x3)
+    for(i in 1:n.sites){
+      for(k in 1:n.years){
+        N.tr[i, k] <- rnbinom(n = 1, mu = lam.tr[i, k], size = th)
+    }}
+
+    # create detection covariate values
+    x1.p <- array(x1[,1], dim = c(n.sites, n.surveys, n.years))
+    x4 <- array(as.numeric(scale(runif(n = n.sites * n.surveys * n.years,
+      -0.5, 0.5), scale = F)), dim = c(n.sites, n.surveys, n.years))
+
+    # average x4 per site-year for example 1
+    x4.m <- apply(x4, c(1, 3), mean, na.rm = F)
+    out1 <- c()
+    for(k in 1:n.years){
+      chunk1 <- x4.m[ , k]
+      chunk2 <- rep(chunk1, n.surveys)
+      out1 <- c(out1, chunk2)
+    }
+    x4.m.arr <- array(out1, dim = c(n.sites, n.surveys, n.years))
+
+    # fill Y.m count array using x4.m for example 1
+    p.tr1 <- plogis(a0 + a1 * x1.p + a4 * x4.m.arr)
+    for (i in 1:n.sites){
+      for (k in 1:n.years){
+        for (j in 1:n.surveys){
+          Y.m[i, j, k] <- rbinom(1, size = N.tr[i, k], prob = p.tr1[i, j, k])
+    }}}
+
+    # fill Y count array using x4 for example 2
+    p.tr2 <- plogis(a0 + a1 * x1.p + a4 * x4)
+    for (i in 1:n.sites){
+      for (k in 1:n.years){
+        for (j in 1:n.surveys){
+          Y[i, j, k] <- rbinom(1, size = N.tr[i, k], prob = p.tr2[i, j, k])
+    }}}
+
+    # format Y.m for data frame output for inla and unmarked
+    Y.m.df <- Y.m[ , , 1]
+    for(i in 2:n.years){
+      y.chunk <- Y.m[ , , i]
+      Y.m.df <- rbind(Y.m.df, y.chunk)
+    }
+
+    # format covariates for data frame output for inla and unmarked
+    x1.df <- rep(x1[ , 1], n.years)
+    x2.df <- rep(x2[ , 1], n.years)
+    x3.df <- rep(x3[1, ], each = n.sites)
+    x1.p.df <- rep(x1.p[ , 1, 1], n.years)
+    x4.df <- c(x4.m)
+
+    # put together data frames for inla and unmarked
+    inla.df <- unmk.df <- data.frame(y1 = Y.m.df[ , 1], y2 = Y.m.df[ , 2],
+    y3 = Y.m.df[ , 3], x1 = x1.df, x2 = x2.df, x3 = x3.df,
+    x1.p = x1.p.df, x4.m = x4.df)
+
+    # return all necessary data for examples 1 and 2
+    return(list(inla.df = inla.df, unmk.df = unmk.df, n.sites = n.sites,
+      n.surveys = n.surveys, n.years = n.years, x1 = x1[ , 1],
+      x2 = x2[ , 1], x3 = x3[1, ], x4 = x4, x4.m = x4.m, x4.m.arr = x4.m.arr,
+      Y = Y, Y.m = Y.m, lam.tr = lam.tr, N.tr = N.tr, x1.p = x1.p[ , 1, 1]
+    ))
+
+  } # end sim.nmix function
R> sim.data <- sim.nmix()
\end{verbatim}

\subsection{Real data}
In addition to simulated data, we also demonstrate the use of \textsf{R-INLA} and \textsf{unmarked} with a real dataset in Example III in Section 5. This dataset comes from a study by \cite{Kery_Royle_Schmid_2005} and is publicly available as part of the \textsf{unmarked} package. The dataset includes mallard duck (\emph{Anas platyrhynchos}) counts, conducted at 239 sites on 2 or 3 occasions during the summer of 2002, as part of a Swiss program that monitors breeding bird abundance (Monitoring H\"{a}ufige Brutv\"{o}gel or Swiss Breeding Bird Survey). In addition to counts, the dataset also includes 2 site-survey covariates related to detection (survey effort and survey date), and 3 site-level covariates related to abundance (route length, route elevation, and forest cover). Full dataset details are given in \cite{Kery_Royle_Schmid_2005}.

\section{Example I}
\subsection{Goals}
In Example I, we demonstrate the use of \textsf{R-INLA} and compare use and performance to similar analyses using \textsf{JAGS} and \textsf{unmarked}. In this first example, the functional forms of \textsf{R-INLA}, \textsf{JAGS}, and \textsf{unmarked} models match the data generating process. Specifically, we used the covariate \texttt{x4.m} to generate the count matrix \texttt{Y.m}, and analyzed the data with models that use \texttt{x4.m} as a covariate. This example was intended to demonstrate the differences and similarities in use, computation time, and estimation results across the three methods when the specified models were the same as the data generating process.

\subsection{Analysis with \textsf{R-INLA}}
We first analyze the simulated data using the \textsf{R-INLA} package. The list returned from the \texttt{sim.nmix()} function includes an object called \texttt{inla.df}. This object has the following structure.  

\begin{verbatim}
R> str(sim.data$inla.df, digits.d = 2)

'data.frame':	648 obs. of  8 variables:
$ y1        : int  2 12 25 3 0 3 1 7 2 8 ...
$ y2        : int  2 22 25 4 1 3 1 11 2 4 ...
$ y3        : int  4 11 28 2 1 2 0 10 2 3 ...
$ x1        : num  0.198 0.353 0.238 0.364 -0.066 ...
$ x2        : num  -0.159 -0.197 -0.484 0.087 0.429 ...
$ x3        : num  -0.5 -0.5 -0.5 -0.5 -0.5 -0.5 -0.5 ...
$ x1.p      : num  0.198 0.353 0.238 0.364 -0.066 ...
$ x4.m      : num  0.148 -0.07 0.206 -0.261 -0.046 ...
R> round(head(sim.data$inla.df), 3)

     y1    y2    y3        x1        x2     x3     x1.p      x4.m
1     2     2     4     0.198    -0.159   -0.5    0.198     0.148
2    12    22    11     0.353    -0.197   -0.5    0.353    -0.070
3    25    25    28     0.238    -0.484   -0.5    0.238     0.206
4     3     4     2     0.364     0.087   -0.5    0.364    -0.261
5     0     1     1    -0.066     0.429   -0.5   -0.066    -0.046
6     3     3     2    -0.356     0.123   -0.5   -0.356    -0.036
\end{verbatim}

This data frame representation of the simulated data has 72 sites $\times$ 9 years = 648 rows. Had there only been one year of data, then the data frame would have 72 rows, one per site. The data frame has three columns (\texttt{y1}, \texttt{y2}, and \texttt{y3}) with count data from the count matrix \texttt{Y.m}, one for each of the three replicate surveys within a given year. Had there been six surveys per year, then there would have been six count columns. The three variables thought to affect abundance are represented in columns 4 through 6. Note that, in this scenario, the first two abundance variables are static across years, so there are 72 unique values in a vector that is stacked 9 times. The third abundance variable, the indicator for year, is a sequence of 9 values, where each value is repeated 72 times. It is centered and scaled in this example. The two variables thought to affect detection probability are represented in columns 7 and 8. The first of these variables has the same values as in column 4, so column 7 is a simple copy of column 4. The second of the two detection variables, shown in column 8, varies per site and year in Example I, so there are 648 unique values in this column. Note that any of the covariates for abundance or detection could have varied by site and year, like \texttt{x4.m}.

We made small modifications to this data frame to prepare data for analysis with \textsf{R-INLA}. In the code that follows, we use the \texttt{inla.mdata()} function to create an object called \texttt{counts.and.count.covs}. The \texttt{counts.and.count.covs} object is essentially a bundle of information related to the abundance component of the model.  Calling the \texttt{str()} function shows that this object is an \textsf{R-INLA} list that includes the three count vectors, passed to the function as a matrix, one vector containing the value of 1, which specifies a global intercept for $\lambda$, and three vectors corresponding to the covariates for $\lambda$. Note that the variable names are standardized by \texttt{inla.mdata()} for computational reasons. 

\begin{verbatim}
R> inla.data <- sim.data$inla.df
R> y.mat <- as.matrix(inla.data[,c("y1", "y2", "y3")])
R> counts.and.count.covs <- inla.mdata(y.mat, 1, inla.data$x1, 
+    inla.data$x2, inla.data$x3)
R> str(counts.and.count.covs)

List of 7
$ Y1: int [1:648] 2 12 25 3 0 3 1 7 2 8 ...
$ Y2: int [1:648] 2 22 25 4 1 3 1 11 2 4 ...
$ Y3: int [1:648] 4 11 28 2 1 2 0 10 2 3 ...
$ X1: num [1:648] 1 1 1 1 1 1 1 1 1 1 ...
$ X2: num [1:648] 0.1983 0.3532 0.2384 0.3636 -0.0661 ...
$ X3: num [1:648] -0.1595 -0.1966 -0.4842 0.0865 0.429 ...
$ X4: num [1:648] -0.5 -0.5 -0.5 -0.5 -0.5 -0.5 -0.5 -0.5 -0.5 -0.5 ...
- attr(*, "class")= chr "inla.mdata"
\end{verbatim}
  
Analysis of N-mixture models with \textsf{R-INLA} is accomplished with a call to the \texttt{inla()} function. The first argument in the \texttt{inla()} call, shown below, is the model formula. On the left side of the formula is the \texttt{counts.and.count.covs} object, which includes the vectors of counts, the global intercept for $\lambda$, and the covariates related to $\lambda$. On the right side of the formula is a 1, to specify a global intercept for $p$, and the two covariates for $p$. Note that a wide range of random effects (exchangeable, spatially or temporally structured) for $p$ could be added to the right side of the formula using the \texttt{f()} syntax \cite{Rue_Riebler_Sorbye_Illian_Simpson_Lindgren_2017}.

The second argument to \texttt{inla()} describes the data, provided here as a list that corresponds with the model formula. Third is the likelihood family, which can take values of \texttt{"nmix"} for a Poisson-binomial mixture and \texttt{"nmixnb"} for a negative binomial-binomial mixture. Run the command \texttt{inla.doc("nmix")} for more information on these likelihood families. The fourth (\texttt{control.fixed}, for detection parameters) and fifth (\texttt{control.family}, for abundance and overdispersion parameters) arguments specify the priors for the two model components. Here, the priors for both abundance and detection parameters are vague normal distributions centered at zero with precision equal to 0.01. The prior for the overdispersion parameter is specified as uniform. Note that a wide variety of other prior distributions are available in \textsf{R-INLA}. At the end of the call are arguments to print the progress of model fitting, and to save information that will enable computation of fitted values. Several other characteristics of the analysis can be modified in a call to \texttt{inla()}, such as whether or not deviance information criterion (DIC), widely applicable information criterion (WAIC), conditional predictive ordinate (CPO), or probability integral transform (PIT) are computed. See \cite{Rue_Riebler_Sorbye_Illian_Simpson_Lindgren_2017} for details.

\begin{verbatim}
R> out.inla.1 <- inla(counts.and.count.covs ~ 1 + x1.p + x4.m,
+    data = list(counts.and.count.covs = counts.and.count.covs,
+      x1.p = inla.data$x1.p, x4.m = inla.data$x4.m),
+    family = "nmixnb",
+    control.fixed = list(mean = 0, mean.intercept = 0, prec = 0.01,
+      prec.intercept = 0.01),
+    control.family = list(hyper = list(theta1 = list(param = c(0, 0.01)),
+      theta2 = list(param = c(0, 0.01)), theta3 = list(param = c(0, 0.01)),
+      theta4 = list(param = c(0, 0.01)), theta5 = list(prior = "flat",
+      param = numeric()))),
+    verbose = TRUE,
+    control.compute=list(config = TRUE))
R> summary(out.inla.1, digits = 3)

Time used (seconds):
Pre-processing    Running inla    Post-processing    Total 
         0.421           5.081              0.342    5.844 

Fixed effects:
              mean     sd  0.025quant  0.5quant 0.975quant
(Intercept)  1.053  0.058       0.938     1.054      1.165
x1.p        -1.996  0.197      -2.385    -1.995     -1.611
x4.m         1.056  0.313       0.440     1.056      1.668

Model hyperparameters:
             mean      sd  0.025quant  0.5quant 0.975quant
beta[1]     2.022   0.034       1.956     2.022      2.090
beta[2]     2.070   0.116       1.839     2.071      2.295
beta[3]    -2.951   0.099      -3.142    -2.953     -2.755
beta[4]     1.142   0.088       0.969     1.142      1.316
overdisp    0.349   0.028       0.296     0.349      0.407
\end{verbatim}
  
\begin{figure}[p]
\includegraphics[width=\linewidth]{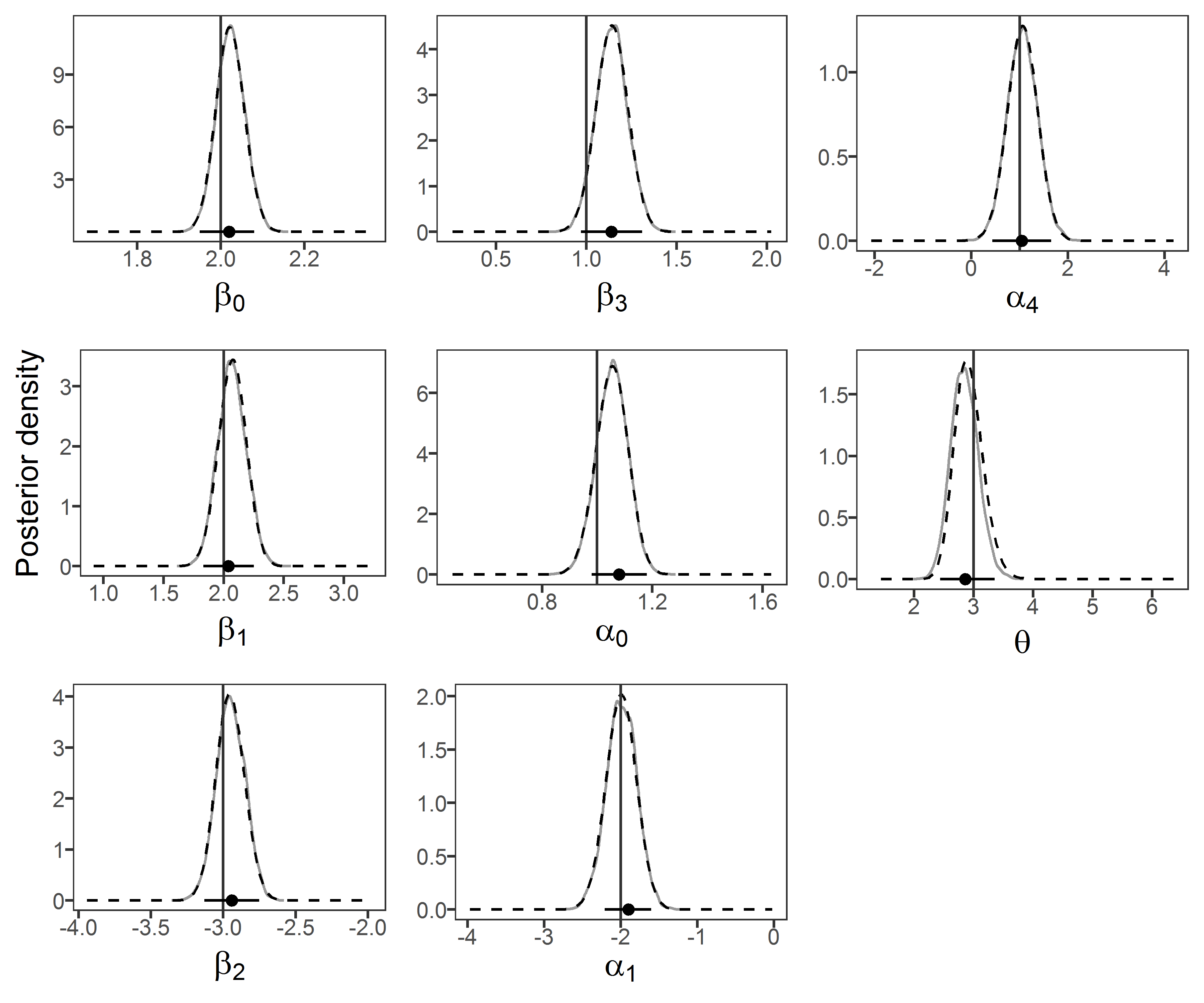}
\caption{Marginal posteriors of model parameters from \textsf{R-INLA} (dashed black lines) and \textsf{JAGS} (solid gray lines), along with Maximum Liklihood estimates (black circles) and 95\% confidence intervals (horizontal black lines) from \textsf{unmarked}.  True input values are represented by vertical black lines.}
\label{fig:fig1}
\end{figure}

Partial output from the \texttt{summary()} function, run on the \texttt{out.inla.1} object, returned from the \texttt{inla()} function, is shown above. The analysis of the model took approximately 6 seconds.  Information on the intercept and covariates related to detection are found under the fixed effects section.  Note that the posterior median parameter estimates related to the detection intercept ($\alpha_0$ labeled as \texttt{(Intercept)}) and covariates ($\alpha_1$ as \texttt{x1.p} and $\alpha_4$ as \texttt{x4.m}) are very close to, and not significantly different from, input parameter values (Fig. \ref{fig:fig1}). Information on abundance and overdispersion parameters are given in the model hyperparameters section. Posterior median estimates for $\beta_0$ (labeled as \texttt{beta[1]}), $\beta_1$ (\texttt{beta[2]}), $\beta_2$ (\texttt{beta[3]}), $\beta_3$ (\texttt{beta[4]}), and $\theta$ (1 / \texttt{overdisp} = 2.87) are also very close to, and not significantly different from, input parameter values (Fig. \ref{fig:fig1}). Density plots of the full marginal posterior distributions for model parameters (Fig. \ref{fig:fig1}) can be viewed using \texttt{plot(out.inla.1)}.

$\lambda_{i,k}$ for each site-year combination in the dataset can be computed using covariate values given in the \texttt{counts.and.count.covs} object, combined with posterior distributions of parameters in the linear predictor of $\lambda_{i,k}$. Posterior distributions for computed $\lambda_{i,k}$ values can be estimated by repeated sampling from the posteriors of hyperparameters, using the \texttt{inla.hyperpar.sample()} function, and repeated solving of the linear predictor. The helper function, \texttt{inla.nmix.lambda.fitted()} produces fitted lambda values as described, using the information contained in the model result output. A call to this function, specifying the model result, estimated posterior sample size, and summary output, is as follows.

\begin{verbatim}
R> out.inla.1.lambda.fits <- inla.nmix.lambda.fitted(result = out.inla.1, 
+    sample.size = 5000, return.posteriors = F)$fitted.summary
R> head(out.inla.1.lambda.fits)

  index mean.lambda sd.lambda  q025.lambda  median.lambda   q975.lambda
1     1     10.3329    0.6623       9.0980        10.3109       11.6683
2     2     15.9003    1.1895      13.6760        15.8649       18.3523
3     3     29.2742    2.2882      25.0410        29.2043       34.0370
4     4      7.0490    0.5250       6.0663         7.0270        8.1227
5     5      1.0547    0.0810       0.9072         1.0509        1.2194
6     6      1.4272    0.1031       1.2389         1.4255        1.6353
\end{verbatim}

The output from this function call is a summary of estimated posteriors for fitted $\lambda_{i,k}$ values. In this example, there are 648 rows. Comparisons of posterior median fitted $\lambda_{i,k}$ with simulated $\lambda_{i,k}$ and $N_{i,k}$ values are shown below.

\begin{verbatim}
R> summary(out.inla.1.lambda.fits$median.lambda)

  Min.   1st Qu.    Median     Mean    3rd Qu.       Max. 
0.6538    3.4020    7.0070  13.5200    16.0300   123.2000
R> summary(c(sim.data$lam.tr))

  Min.   1st Qu.    Median     Mean    3rd Qu.       Max. 
0.6834    3.3050    6.8190  13.0300    15.9500   111.3000 
R> cor(out.inla.1.lambda.fits$median.lambda, c(sim.data$lam.tr))

[1] 0.9986
R> sum(out.inla.1.lambda.fits$median.lambda)

[1] 8758.542
R> sum(c(sim.data$lam.tr))

[1] 8444.975
R> sum(c(sim.data$N.tr))

[1] 8960
\end{verbatim}
  
\subsection{Analysis with \textsf{JAGS}}
Next, we analyzed the same simulated dataset using \textsf{JAGS}, via the \textsf{runjags} package. As for the \textsf{R-INLA} analysis, we specified a negative binomial distribution for abundance, vague normal priors for the intercepts and the global effects of the covariates of $\lambda$ and $p$, and a flat prior for the overdispersion parameter.  The \textsf{JAGS} model statement, where the distributions and likelihood function are specified, is shown below for comparison with the arguments to \texttt{inla()}.

\begin{verbatim}
R> jags.model.string <- "
+    model {
+      a0 ~ dnorm(0, 0.01)
+      a1 ~ dnorm(0, 0.01)
+      a4 ~ dnorm(0, 0.01)
+      b0 ~ dnorm(0, 0.01)
+      b1 ~ dnorm(0, 0.01)
+      b2 ~ dnorm(0, 0.01)
+      b3 ~ dnorm(0, 0.01)
+      th ~ dunif(0, 5)
+    for (k in 1:n.years){
+      for (i in 1:n.sites){
+        N[i, k] ~ dnegbin(prob[i, k], th)
+        prob[i, k] <- th / (th + lambda[i, k])
+        log(lambda[i, k]) <- b0 + (b1 * x1[i]) + (b2 * x2[i]) + (b3 * x3[k])
+        for (j in 1:n.surveys){
+          Y.m[i, j, k] ~ dbin(p[i,j,k], N[i,k])
+          p[i, j, k] <- exp(lp[i,j,k]) / (1 + exp(lp[i,j,k]))
+          lp[i, j, k] <- a0 + (a1 * x1.p[i]) + (a4 * x4.m[i, k])
+    }}}}
+  "
\end{verbatim}

After specifying the \textsf{JAGS} model, we define the parameters to be monitored during the MCMC simulations, bundle numerous values and vectors from the \texttt{sim.data} object, and create a function for drawing random initial values for the model parameters. These steps are included in the code supplement, but are not shown here. Finally, we set the run parameters, such as the number of chains and iterations, and start the MCMC process. Run parameters were chosen such that MCMC diagnostics indicated converged chains (potential scale reduction factors $\leq$ 1.05) and reasonably robust posterior distributions (effective sample sizes $\geq$ 3000). Note that the recommended number of effective samples for particularly robust inference is closer to 6000 \cite{Gong_Flegal_2016}. Thus, MCMC processing times reported here could be considered optimistic estimates. The MCMC simulation is initiated with a call to \texttt{run.jags()}. Partial output from the simulation, related to parameter estimates, is shown below.

\begin{verbatim}
R> out.jags.1 <- run.jags(model = jags.model.string, data = jags.data, 
+    monitor = params, n.chains = 3, inits = inits, burnin = 3000, 
+    adapt = 3000, sample = 6000, thin = 10, modules = "glm on", 
+    method = "parallel")
R> round(summary(out.jags.1), 3)[ , c(1:5, 9, 11)]

            Lower95 Median Upper95   Mean    SD  SSeff psrf
a0            0.941  1.053   1.161  1.053 0.057   3155    1
a1           -2.380 -1.990  -1.618 -1.994 0.195   2954    1
a4            0.434  1.053   1.661  1.052 0.314   5695    1
b0            1.956  2.024   2.089  2.024 0.034   7065    1
b1            1.851  2.070   2.301  2.071 0.115   6028    1
b2           -3.142 -2.946  -2.755 -2.947 0.099  17539    1
b3            0.969  1.142   1.315  1.142 0.089  18000    1
th            2.401  2.840   3.295  2.850 0.230  18000    1
\end{verbatim}

Similar to the \textsf{R-INLA} analysis, median parameter estimates from the \textsf{JAGS} model were close to, and not significantly different from, the input values used to generate the data (Fig. \ref{fig:fig1}). The potential scale reduction factor for all variables was $\geq$ 1.05, and the effective sample size for all variables was approximately 3000 or greater. The simulation ran in parallel on 3 virtual cores, 1 MCMC chain per core, and took approximately 2960 seconds.

\subsection{Analysis with \textsf{unmarked}}
Lastly, we prepare the simulated data for the \textsf{unmarked} analysis, which involved slight modification of the \texttt{unmk.df} object created using the \texttt{sim.nmix()} function. As with the \textsf{JAGS} analysis, these steps are included in the code supplement, but are not illustrated here.

The \textsf{unmarked} analysis is run by a call to the \texttt{pcount()} function. The first argument in the call to \texttt{pcount()} is the model formula, which specifies the covariates for detection, and then the covariates for abundance. This is followed by an argument identifying the unmarked data object, and the form of the mixture model, negative binomial-binomial in this case.

\begin{verbatim}
R> out.unmk.1 <- pcount(~ 1 + x1.p + x4.m ~ 1 + x1 + x2 + x3,
+    data = unmk.data, mixture = "NB")
R> summary(out.unmk.1)

Abundance (log-scale):
            Estimate       SE       z     P(>|z|)
(Intercept)     2.02   0.0321    62.9    0.00e+00
x1              2.04   0.1071    19.0    1.69e-80
x2             -2.94   0.0982   -29.9    5.36e-20
x3              1.14   0.0882    13.0    2.26e-38

Detection (logit-scale):
            Estimate       SE       z     P(>|z|)
(Intercept)     1.08   0.0507   21.27   2.29e-100
x1.p           -1.90   0.1576  -12.08    1.30e-33
x4.m            1.04   0.3102    3.35    8.02e-04

Dispersion (log-scale):
            Estimate       SE       z     P(>|z|)
                1.05   0.0804    13.1    4.91e-39
\end{verbatim}

Maximum likelihood estimates for model parameters from \textsf{unmarked} were also close to, and not significantly different from, input values (Fig. \ref{fig:fig1}). Note that the dispersion estimate, after exponentiation, was 2.86. The \textsf{unmarked} estimates were produced in approximately 86 seconds.

\subsection{Example I summary}
Example I demonstrated basic use of \textsf{R-INLA} to analyze N-mixture models and highlighted similarities and differences between it and two other commonly used approaches. In demonstrating the use of \textsf{R-INLA}, we showed that the input data format is not too complicated, and that the formatting process can be accomplished with a few lines of code. Similarly, model specification uses a straightforward extension of the standard syntax in \textsf{R}, where the counts and covariates for $\lambda$ are specified through an \textsf{R-INLA} object included on the left side of the formula, and fixed covariates and random effects for $p$ are specified on the right side of the formula. The data format and model specification syntax of \textsf{R-INLA} is not too different from \textsf{unmarked}, whereas those of both packages are considerably different from \textsf{JAGS} and other MCMC software, such as \textsf{OpenBUGS}, \textsf{WinBUGS}, and \textsf{Stan}.

Regarding performance, \textsf{R-INLA}, \textsf{JAGS}, and \textsf{unmarked} all successfully extracted simulation input values. Fig. \ref{fig:fig1} shows marginal posterior distributions produced by \textsf{R-INLA} and \textsf{JAGS}, and estimates and 95\% confidence intervals from \textsf{unmarked}. These results derive from data from one random manifestation of the input values. Thus, we do not expect the posterior distributions for the estimates to be centered at the input values, which would be expected if the simulation was repeated many times. However, we do expect the input values to fall somewhere within the posterior distributions and 95\% confidence limits, which is what occured here. Fig. \ref{fig:fig1} shows that, for similarly specified models, \textsf{R-INLA} (dashed black lines) and \textsf{JAGS} (solid gray lines) yielded practically identical marginal posterior distributions for model parameters. Fig. \ref{fig:fig1} also illustrates the general agreement between the credible intervals associated with \textsf{R-INLA} and \textsf{JAGS} and the confidence intervals associated with \textsf{unmarked}.

Where \textsf{R-INLA}, \textsf{JAGS}, and \textsf{unmarked} differed substantially was in computing time. In this example, \textsf{R-INLA} took 6 seconds, \textsf{JAGS} took 2960 seconds, and \textsf{unmarked} took 86 seconds to produce results. Thus, \textsf{R-INLA} was approximately 500 times faster than \textsf{JAGS} and 10 times faster than \textsf{unmarked}. This was the case despite the fact that \textsf{unmarked} produced ML estimates and the \textsf{JAGS} analysis was run in parallel with each of three MCMC chains simulated on a separate virtual computing core. If parallel computing had not been used with \textsf{JAGS}, processing the \textsf{JAGS} model would have taken approximately twice as long. If MCMC simulations were run until effective sample sizes of 6000 were reached, processing time would have doubled again.

In sum, when compared to other tools, \textsf{R-INLA} is relatively easy to implement and produces accurate estimates of Bayesian posteriors very quickly. Its utility depends on the degree to which the data generating process can be captured accurately in model specification. However, as mentioned above, certain N-mixture models can not be specified using \textsf{R-INLA}. For the data in Example I, the count matrix was produced using a detection covariate that was averaged to the site-year level. This averaged covariate was subsequently specified in the model. But what happens when the site-survey-year covariate is an important component of the data generating process, and it can't be entered into the model in this form? This is the question explored in Example II.

\section{Example II}
\subsection{Goals}
In Example II, we show the consequences of not being able to specify a site-survey-year covariate for detection, under a range of conditions. We conducted a Monte Carlo experiment where, for each iteration, the count matrix for the analysis, \texttt{Y}, was generated with the \texttt{sim.nmix()} function using the site-survey-year covariate \texttt{x4}. The count data were then analyzed with two \textsf{JAGS} models. The first model incorporated the site-survey-year \texttt{x4} covariate. The second model incorporated the averaged site-year \texttt{x4.m}, instead. For each iteration, we randomly varied the size of $\alpha_4$ when generating the simulated data. We expected that the simpler model, with \texttt{x4.m}, would yield biased estimates when the magnitude of $\alpha_4$ was relatively large, and unbiased estimates when the magnitude of $\alpha_4$ was relatively small. All computing code related to Example II is given in the supplemental code file.

\subsection{Analysis with \textsf{JAGS}}
\begin{figure}
\includegraphics[width=\linewidth]{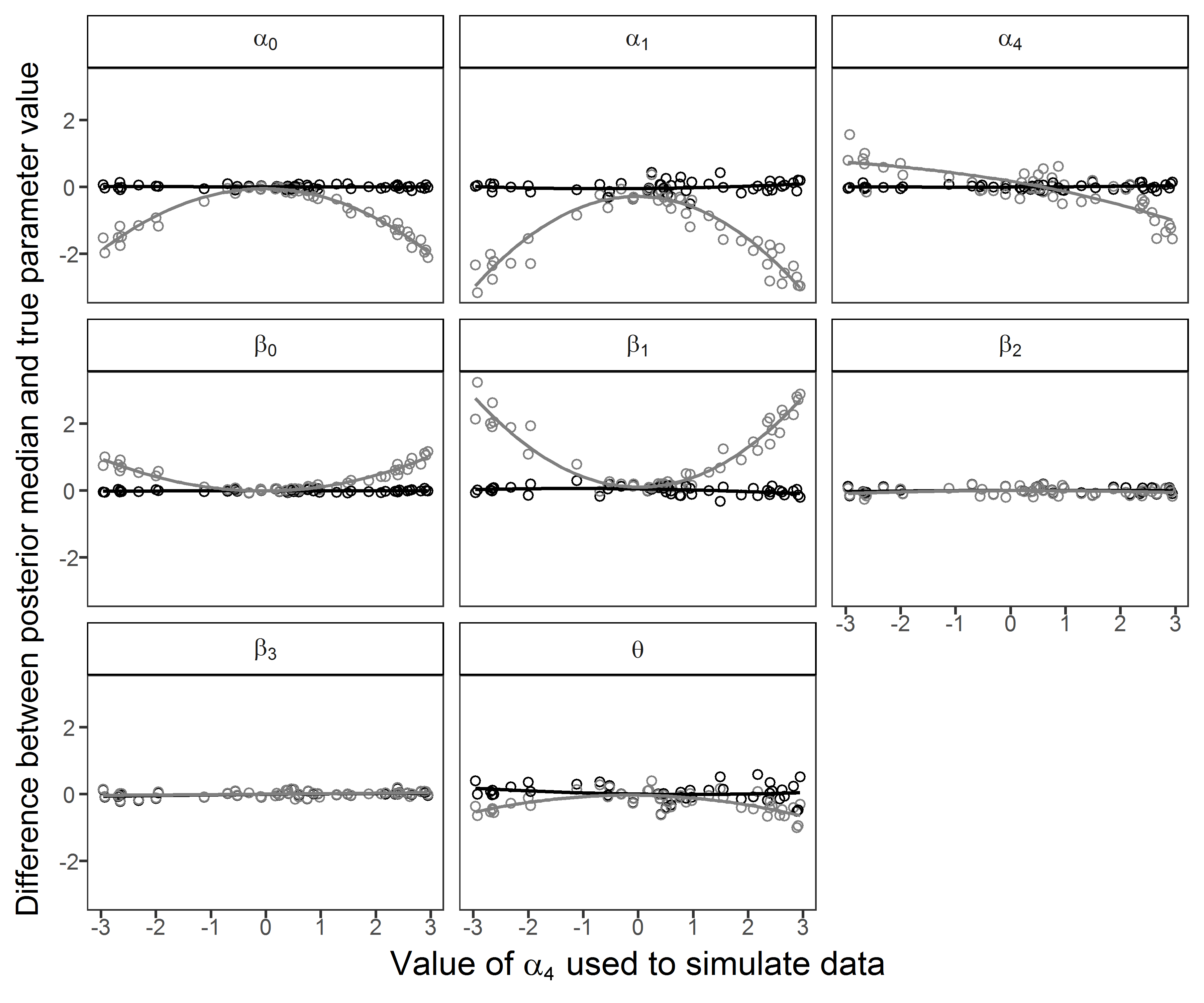}
\caption{Differences between posterior median parameter values and true input parameter values as a function of the $\alpha_4$ value used to simulate data. Black circles and lines are from the model with the site-survey-year covariate, \texttt{x4}, and gray circles and lines are from the model with an averaged site-year covariate, \texttt{x4.m}. Parameter name is given in the strip across the top of each panel.}
\label{fig:fig2}
\end{figure}

Parameter values entered into \texttt{sim.nmix()}, other than those for $\alpha_4$, were the same as those used in Example I.  Similarly, the \textsf{JAGS} model specification, other than parts associated with $\alpha_4$, was the same as that used in Example I. Given the long processing time associated with \textsf{JAGS} models in Example I, we only ran and saved 1000 MCMC simulations (no thinning, after 500 adaptive and 100 burn-in iterations) during each of the 50 Monte Carlo runs in Example II. This number is not sufficient for drawing inference from marginal posteriors, but was sufficient for looking at qualitative patterns in posterior medians. For each of these runs, a value for $\alpha_4$ was drawn from a uniform distribution that ranged from -3 to 3. Parameter bias was represented for each model parameter as the difference between the simulation input and the posterior median estimated value. The results of the simulations are depicted in Fig. \ref{fig:fig2}.

\subsection{Example II summary}
Even with as few as 50 Monte Carlo runs, it was apparent that biases in parameter estimates increased with the magnitude of $\alpha_4$ (Fig. \ref{fig:fig2}). When the magnitude of $\alpha_4$ was small, with an absolute value less than 1, the bias was negligible. When the magnitude of $\alpha_4$ was large, with an absolute value greater than 2, the bias was considerable (Fig. \ref{fig:fig2}). When interpreting the effect size, bear in mind that \texttt{x4} ranged from -0.5 to 0.5.

\section{Example III}
\subsection{Goals}
In Example III, we explore the performance of \textsf{R-INLA} using real data, a publicly available dataset of mallard duck counts from Switzerland during 2002. By employing real data, we hoped to evaluate (\textit{i}) the performance of \textsf{R-INLA} using data that were not predictable by design and (\textit{ii}) the practical consequences of not being able to specify site-survey covariates in \textsf{R-INLA}. The dataset is available as a demonstration dataset in \textsf{unmarked}, so we compared the performance of \textsf{R-INLA} with that of \textsf{unmarked}, using the analysis settings and model structure described in \textsf{unmarked} documentation.

\subsection{Analysis with \textsf{R-INLA}}
The mallard data is provided in the \textsf{unmarked} package as a list with three components: a matrix of counts (\texttt{mallard.y}), a list of matrices of detection covariates (\texttt{mallard.obs}), and a data frame of abundance covariates (\texttt{mallard.site}). Data in \textsf{unmarked} are organized in structures called unmarked frames, which are viewed as a data frame when printed.

\begin{verbatim}
R> data(mallard)
R> mallard.umf <- unmarkedFramePCount(y = mallard.y, siteCovs = 
+    mallard.site, obsCovs = mallard.obs)
R> mallard.umf[1:6, ]

Data frame representation of unmarkedFrame object.
     y1   y2   y3    elev length forest  ivel1  ivel2  ivel3  
1     0    0    0  -1.173  0.801 -1.156 -0.506 -0.506 -0.506 
2     0    0    0  -1.127  0.115 -0.501 -0.934 -0.991 -1.162 
3     3    2    1  -0.198 -0.479 -0.101 -1.136 -1.339 -1.610 
4     0    0    0  -0.105  0.315  0.008 -0.819 -0.927 -1.197 
5     3    0    3  -1.034 -1.102 -1.193  0.638  0.880  1.042 
6     0    0    0  -0.848  0.741  0.917 -1.329 -1.042 -0.899 
...
    date1  date2  date3
1  -1.761  0.310  1.381
2  -2.904 -1.047  0.596
3  -1.690 -0.476  1.453
4  -2.190 -0.690  1.239
5  -1.833  0.167  1.381
6  -2.619  0.167  1.381
\end{verbatim}

As discussed above, it is not possible to take advantage of survey-level covariates when analyzing N-mixture models with \textsf{R-INLA}.  So, before analysis with \textsf{R-INLA}, we averaged the survey-level variables, \texttt{ivel} and \texttt{date}, per site using the \texttt{rowMeans()} function.

\begin{verbatim}
R> length <- mallard.site[ , "length"]
R> elev <- mallard.site[ , "elev"]
R> forest <- mallard.site[ , "forest"]
R> mean.ivel <- rowMeans(mallard.obs$ivel, na.rm = T) 
R> mean.ivel[is.na(mallard.ivel)] <- mean(mallard.ivel, na.rm = T)
R> mean.date <- rowMeans(mallard.obs$date, na.rm = T) 
R> mean.date.sq <- mean.date^2
R> mallard.inla.df <- data.frame(y1 = mallard.y[ , "y.1"], 
+    y2 = mallard.y[ , "y.2"], y3 = mallard.y[ , "y.3"], 
+    length, elev, forest, mean.ivel, mean.date, mean.date.sq)
R> round(head(mallard.inla.df), 3)

  y1 y2 y3 length   elev forest mean.ivel mean.date mean.date.sq
1  0  0  0  0.801 -1.173 -1.156    -0.506    -0.023        0.001
2  0  0  0  0.115 -1.127 -0.501    -1.029    -1.118        1.251
3  3  2  1 -0.479 -0.198 -0.101    -1.362    -0.238        0.056
4  0  0  0  0.315 -0.105  0.008    -0.981    -0.547        0.299
5  3  0  3 -1.102 -1.034 -1.193     0.853    -0.095        0.009
6  0  0  0  0.741 -0.848  0.917    -1.090    -0.357        0.127
\end{verbatim}

The data are now in a format that can be analyzed readily using \textsf{R-INLA}. The data frame has 239 sites $\times$ 1 year = 239 rows, one column for each replicate count, and one column for each detection and abundance covariate. Once in this form, it is easy to create an \texttt{inla.mdata()} object and run the analysis. In preparing \texttt{counts.and.count.covs}, we specify an intercept and effects of transect length (\texttt{length}), elevation (\texttt{elev}), and forest cover (\texttt{forest}) on abundance. In the \texttt{model} argument to \texttt{inla()}, we specify an intercept and effects of survey intensity (\texttt{ivel}) and survey date (\texttt{date}) for detection. As before, the data argument is a list that corresponds with the model formula. The family argument specifies a negative binomial-binomial mixture. The priors for intercepts and covariates are specified as vague normal distributions, and that for the overdispersion parameter as a uniform distribution. 

\begin{verbatim}
R> counts.and.count.covs <- inla.mdata(mallard.y, 1, length, elev, forest)
R> out.inla.2 <- inla(counts.and.count.covs ~ 1 + mean.ivel +
+    mean.date + mean.date.sq, 
+    data = list(counts.and.count.covs = counts.and.count.covs, 
+      mean.ivel = mallard.inla.df$mean.ivel, mean.date = 
+      mallard.inla.df$mean.date, mean.date.sq = mallard.inla.df$mean.date.sq), 
+    family = "nmixnb", 
+    control.fixed = list(mean = 0, mean.intercept = 0, prec = 0.01, 
+      prec.intercept = 0.01),
+    control.family = list(hyper = list(theta1 = list(param = c(0, 0.01)),
+      theta2 = list(param = c(0, 0.01)), theta3 = list(param = c(0, 0.01)),
+      theta4 = list(param = c(0, 0.01)), theta5 = list(prior = "flat",
+      param = numeric()))))
R> summary(out.inla.2, digits = 3)
\end{verbatim}

A portion of the summary for \texttt{out.inla.2} is shown below. Note that posterior summaries described in the fixed effects section pertain to the intercept and covariates of $p$. In the hyperparameters section, \texttt{beta[1]}, \texttt{beta[2]}, \texttt{beta[3]}, and \texttt{beta[4]} identify posterior summaries for the $\lambda$ intercept, and transect length, elevation, and forest cover effects.

\begin{verbatim}
Fixed effects:
               mean    sd 0.025quant  0.5quant  0.975quant
(Intercept)  -0.397 0.383     -1.170    -0.389       0.335
mean.ivel     0.039 0.212     -0.378     0.039       0.455
mean.date    -1.044 0.433     -1.923    -1.036      -0.195
mean.date.sq -0.318 0.304     -0.962    -0.301       0.233

Model hyperparameters:
             mean     sd  0.025quant  0.5quant  0.975quant
beta[1]    -1.412  0.296      -1.966    -1.424      -0.801 
beta[2]    -0.290  0.190      -0.664    -0.291       0.086
beta[3]    -0.998  0.318      -1.595    -1.011      -0.341
beta[4]    -0.771  0.203      -1.178    -0.767      -0.382
overdisp    1.228  0.264       0.799     1.194       1.837
\end{verbatim}

\subsection{Analysis with \textsf{unmarked}}
The \textsf{unmarked} frame, \texttt{mallard.umf}, created above, can be used directly by the \texttt{pcount()} function in \textsf{unmarked}. The data and model structure described in the \texttt{pcount()} function below is similar to that used above in the \textsf{R-INLA} analysis, except for one key difference: here, \texttt{ivel} and \texttt{date} are site-survey level variables instead of the site-level means used in the \textsf{R-INLA} analysis.

\begin{verbatim}
R> out.unmk.2 <- pcount(~ ivel+ date + I(date^2) ~ length + elev + forest,
+    mixture = "NB", mallard.umf)
R> summary(out.unmk.2)

Abundance (log-scale):
              Estimate         SE           z     P(>|z|)
(Intercept)     -1.786      0.281      -6.350    2.15e-10
length          -0.186      0.214      -0.868    3.86e-01
elev            -1.372      0.293      -4.690    2.73e-06
forest          -0.685      0.216      -3.166    1.54e-03

Detection (logit-scale):
              Estimate         SE           z     P(>|z|)
(Intercept)     -0.028      0.285      -0.099       0.921
ivel             0.174      0.227       0.766       0.444
date            -0.313      0.147      -2.132       0.033
I(date^2)       -0.005      0.081      -0.059       0.953

Dispersion (log-scale):
              Estimate         SE           z     P(>|z|)
                -0.695      0.364       -1.91       0.056
\end{verbatim}

\begin{figure}[p]
\includegraphics[width=5in]{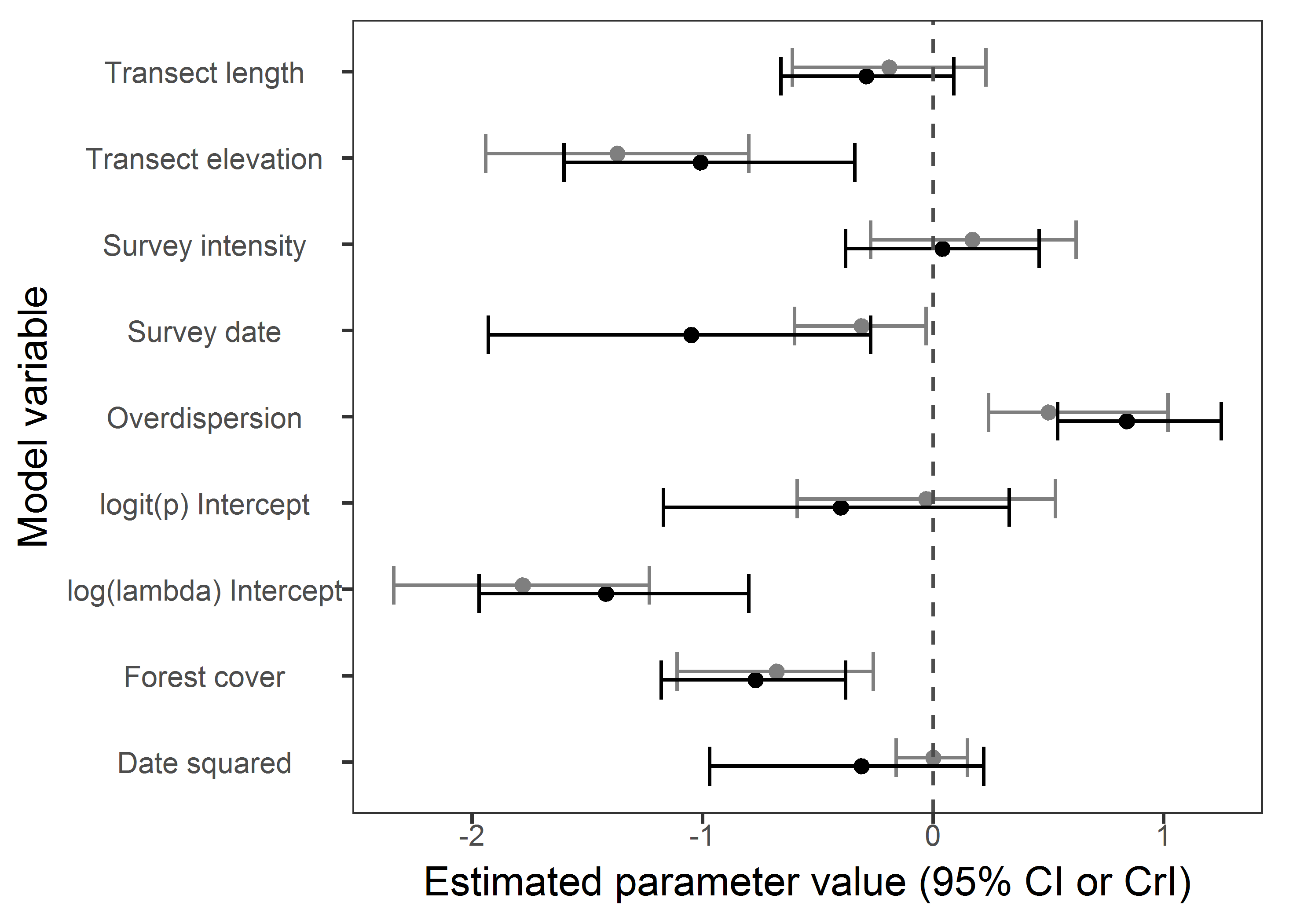}
\caption{Parameter estimates and 95\% confidence intervals from \textsf{unmarked} (gray circles and lines) and posterior medians and 95\% credible intervals from \textsf{R-INLA} (black circles and lines) from an N-mixture model analysis of mallard duck abundance. Model parameters are identified by their associated variable names listed on the vertical axis.  The \textsf{unmarked} model included site-survey covariates for survey intensity and survey date, while the \textsf{R-INLA} model included site-averaged versions. A value of zero (no effect) is depicted by the vertical dashed gray line.}
\label{fig:fig3}
\end{figure}

\subsection{Example III summary}
Comparing the results, we see that the 95\% credible intervals for parameter estimates from the \textsf{R-INLA} analysis overlapped broadly with the 95\% confidence intervals from the \textsf{unmarked} analysis, so parameter estimates were not significantly different from one another (Fig. \ref{fig:fig3}). Regardless of technique, the same set of parameters had estimates significantly different from zero (Fig. \ref{fig:fig3}), and significant effects were of the same magnitude and direction in both analyses. Using both techniques, detection decreased as the season progressed, and abundance decreased with increasing forest cover and elevation. Parameter estimates and biological conclusions were similar despite the fact that site-survey detection covariates were used for \textsf{unmarked} and site-averaged detection covariates were used for \textsf{R-INLA}. Note that \textsf{unmarked} estimated moderate effects of detection covariates which, according to the results in Example II, would indicate that parameter estimates from the \textsf{R-INLA} analysis were not substantially biased. These conclusions may have been different given a different dataset, where detection covariates had very strong effects or were not otherwise controlled by survey design.

\section{Discussion}
The purpose of this work was to detail the use of the \textsf{R-INLA} package \cite{Rue_Riebler_Sorbye_Illian_Simpson_Lindgren_2017} to analyze N-mixture models and to compare analyses using \textsf{R-INLA} to two other common approaches: \textsf{JAGS} \cite{plummer2003jags,Lunn_Jackson_Best_Thomas_Spiegelhalter_2012}, via the \textsf{runjags} package \cite{Denwood_2016}, which employs MCMC methods and allows Bayesian inference, and the \textsf{unmarked} package \cite{Fiske_Chandler_2011}, which uses maximum likelihood and allows frequentist inference. While we selected \textsf{JAGS} as the representative MCMC approach, we expect that our conclusions would be qualitatively similar for other MCMC software, such as \textsf{OpenBUGS}, \textsf{WinBUGS}, or \textsf{Stan}. We are not aware of other commonly-used software for analyzing N-mixture models in a maximum likelihood framework, besides \textsf{unmarked}.

Comparisons showed that \textsf{R-INLA} can be a complementary tool in the wildlife biologist's analytical tool kit. Strengths of \textsf{R-INLA} include Bayesian inference, based on highly accurate approximations of posterior distributions, which were derived roughly 500 times faster than MCMC methods, where models are specified using a syntax that should be familiar to R users, and where data are formatted in a straightforward way with relatively few lines of code. The straightforward model syntax and data format could help lower  barriers to adoption of N-mixture models for biologists who are not committed to learning  \textsf{BUGS} or \textsf{Stan} syntax. The substantial decrease in computation time should facilitate use of a wider variety of model and variable selection techniques (e.g., cross validation and model averaging), ones that are not commonly used in an MCMC context due to practical issues related to computing time \cite{Kery_Schaub_2011}.

Limitations of \textsf{R-INLA} are mainly related to the more restricted set of N-mixture models that can be specified. Of the approaches described here, ones that use MCMC allow users ultimate flexibility in specifying models. For example, with \textsf{JAGS}, site-survey covariates for detection are possible, multiple types of mixed distributions are available \cite{Joseph_Elkin_Martin_Possingham_2009,Martin_Royle_Mackenzie_Edwards_Kery_Gardner_2011}, and a variety of random effects can be specified for both $\lambda$ and $p$ \cite{Kery_Schaub_2011}. In comparison, the current version of \textsf{R-INLA} does not handle site-survey covariates, employs only Poisson-binomial and negative binomial-binomial mixtures, and handles random effects for $p$ only. A practical consequence of the random effects limitation is that, while site and site-year posteriors for $\lambda$ can be estimated using \textsf{R-INLA}, site and site-year posteriors for $N$ are not currently available (see Appendix). In cases where site-survey covariates are particularly important, and not otherwise controlled by survey design, where different mixed distributions are required, or where random effects associated with $\lambda$ are needed, an MCMC approach appears to be most appropriate (Fig. \ref{fig:fig2}).

When compared to \textsf{unmarked}, the \textsf{R-INLA} approach is similar in regards to familiar model syntax and data format. The approaches are also similar in that both yield results much faster than MCMC, enabling a richer set of options in terms of model and variable selection. The two approaches differ in that \textsf{R-INLA} is approximately 10 times faster than \textsf{unmarked}, likely due to the different method used to compute model likelihoods (see Appendix). They also differ in that \textsf{unmarked} can accommodate site-survey covariates, whereas \textsf{R-INLA} does not, and that \textsf{R-INLA} can accomodate random effects for $p$, whereas \textsf{unmarked} does not. In cases where both computing speed and specification of site-survey covariates are critical, \textsf{unnmarked} appears to an appropriate tool.

In conclusion, \textsf{R-INLA}, \textsf{JAGS} (and \textsf{WinBUGS}, \textsf{OpenBUGS}, and \textsf{Stan}), and \textsf{unmarked} all allow users to analyze N-mixture models for estimating wildlife abundance while accounting for imperfect detection. Each method has its strengths and limitations. \textsf{R-INLA} appears to be an attractive option when survey-level covariates are not essential, familiar model syntax and data format are desired, Bayesian inference is preferred, and fast computing time is required.

\section*{Acknowledgments}
We thank C. Burkhalter, T. Onkelinx, U. Halekoh, E. Pebesma, and an anonymous reviewer for commenting on previous drafts of this manuscript.

\bibliography{bibliog}
\bibliographystyle{ieeetr}

\section*{Appendix}
\subsection*{A. Posterior probability for $N$}
Currently, it is not possible to extract posteriors for $N$ when analyzing N-mixture models using \textsf{R-INLA}. This functionality, which would utilize output from the \texttt{inla.posterior.sample()} function, could be available in future versions based on the following logic. Assume the Poisson for $N$, such that $\textnormal{Prob} (N | \lambda) = p_0(N; \lambda)$, and

$$\textnormal{Prob} (y_1, ..., y_m|N) = \sum \bigg[ \prod_{i=1}^{m} \textnormal{Bin} (y_i; N, p) \bigg] p_0(N; \lambda),$$

where $y_1, ..., y_m = \Upsilon$, and $N \geq \textnormal{max} (y_1, ..., y_m)$. If we have samples from the posterior of $(p, \lambda) | \Upsilon$, we can compute the posterior marginal of $N | \Upsilon$ as follows. If $(p, \lambda)$ is fixed, then

$$\textnormal{Prob} (N | \Upsilon) \propto \bigg[ \prod_{i=1}^{m} \textnormal{Bin} (y_i; N, p) \bigg] p_0 (N; \lambda),$$

and this expression is evaluated for $N = \textnormal{max}(y_1, ..., y_m)$, ..., and renormalized. We can integrate out $(p, \lambda) | \Upsilon$ using samples from the posteriors, as

$$\textnormal{Prob} (N | \Upsilon) = \frac{1}{M}  \sum_{j=1}^{M} \frac{1}{Z(p_j,\lambda_j)}  \bigg[ \prod_{i=1}^{M} \textnormal{Bin} (y_i; N, p_j) \bigg] p_0 (N; \lambda_j),$$

for $M$ samples $(p_1, \lambda_1), ..., (p_M, \lambda_M)$ from the posterior of $(p, \lambda) | \Upsilon$. That is, we average the probability for each $N$, renormalize, and normalize for each sample by computing $Z (p, \lambda)$.

\subsection*{B. Recursive computations of the 'nmix' likelihood}
The likelihood for the simplest case is

\begin{displaymath}
\text{Prob}(y) = \sum_{N = y}^{\infty}
\text{Pois}(N ; \lambda) \;\times\; \text{Bin}(y;  N, p)
\end{displaymath}

where $\text{Pois}(N; \lambda)$ is the density for the Poisson distribution with mean $\lambda$, $\lambda^{N}\exp(-\lambda)/N!$, and $\text{Bin}(y; N, p)$ is the density for the binomial distribution with $N$ trials and probability $p$, ${N \choose y} p^{y}(1-p)^{N-p}$. Although the likelihood can be computed directly when replacing the infinite limit with a finite value, we will demonstrate here that we can easily evaluate it using a recursive algorithm that is both faster and more numerical stable. The same idea is also applicable to the negative binomial case, and the case where we have replicated observations of the same $N$. We leave it to the reader to derive these straight forward extensions.

The key observation is that both the Poisson and the binomial distribution can be evaluated recursively in $N$,

\begin{displaymath}
\text{Pois}(N; \lambda) = \text{Pois}(N-1; \lambda) \frac{\lambda}{N}
\end{displaymath}

and

\begin{displaymath}
\text{Bin}(y; N, p) = \text{Bin}(y; N-1, p) \frac{N}{N-y}(1-p),
\end{displaymath}

and then also for the Poisson-binomial product

\begin{displaymath}
\text{Pois}(N ; \lambda) \; \text{Bin}(y; N, p)
=
\text{Pois}(N-1; \lambda) \; \text{Bin}(y; N-1, p)
\frac{\lambda}{N-y}(1-p).
\end{displaymath}

If we define $f_i = \lambda(1-p)/i$ for $i=1, 2, \ldots$, we can make use of this recursive form to express the likelihood with a finite upper limit as

\begin{eqnarray}
\text{Prob}(y) &=& \sum_{N = y}^{N_{\text{max}}}
\text{Pois}(N ; \lambda)\;
\text{Bin}(y; N, p) \nonumber\\
&=& \text{Pois}(y; \lambda)\; \text{Bin}(y; y, p)
\Big\{ 1 + f_1 + f_1f_2 +
\ldots
+f_1\cdots f_{N_\text{max}}
\Big\} \nonumber\\
&=& \text{Pois}(y; \lambda)\; \text{Bin}(y; y, p)
\Big\{ 1 + f_1(1+f_2(1+f_3(1+ \dots)))\Big\}\nonumber
\end{eqnarray}

The log-likelihood can then be evaluated using the following simple \textsf{R} code.

\begin{verbatim}
R> fac <- 1; ff <- lambda * (1-p)
R> for(i in (N.max - y):1) fac <- 1 + fac * ff / i
R> log.L <- dpois(y, lambda, log = TRUE) +
+    dbinom(y, y, p, log = TRUE) + log(fac)
\end{verbatim}

Since this evaluation is recursive in decreasing $N$, we have to choose the upper limit $N_\text{max}$ in advance, for example as an integer larger than $y$ so that $\frac{\lambda (1-p)}{N_\text{max}-y}$ is small. Note that we are computing \texttt{fac} starting with the smallest contributions, which are more numerically stable.

\end{document}